\documentclass[twocolumn,showpacs,preprintnumbers,amsmath,amssymb]{revtex4}

\usepackage{graphicx}
\usepackage{dcolumn}
\usepackage{bm}

\begin{document} 

\preprint{}
\input{epsf.tex}

\title{Polarization-Mixing in Optical Lattices with Uniaxial Anisotropy}
\author{Hashem Zoubi and Helmut Ritsch}

\affiliation{Institut f\"{u}r Theoretische Physik, Universit\"{a}t Innsbruck, Technikerstrasse 25, A-6020 Innsbruck, Austria}  

\date{04 July, 2008}

\begin{abstract}
We study optical excitations of an ultracold gas in an optical lattice in the Mott insulator case and with anisotropic properties. The anisotropy is induced by an oriented transition dipole generated by optical pumping and external static fields, and is of the type of uniaxial crystals. Within a cavity, the electronic excitations and the cavity photons of both TE and TM polarizations are coherently mixed to form two polariton branches. Also a photonic branch appears which decouples from the electronic transitions. The photon polarization mixing can be observed by linear optical spectra. For example, for an incident field of TE polarization we obtain TE and TM transmitted and reflected fields. Also we calculate the phase shift in the TE and TM transmitted and reflected fields relative to the incident field phase, which lead to polarization rotations.
\end{abstract}


\maketitle

\section{Introduction}

An optical lattice is produced by pairs of counter propagating laser beams, which introduce standing waves of lattice constant of half wave length, $a=\lambda/2$ \cite{Zoller}. The laser beams have a given wave length, intensity, and polarization, and where they are off resonance to the atomic internal transitions. A boson gas of ultracold atoms loaded into the optical lattice can be described by Bose-Hubbard model, and a quantum phase transition from the superfluid into the Mott insulator phase occurs by changing the laser intensity \cite{Jaksch,Bloch}. Here we consider the Mott insulator phase with one atom per lattice site, and we concentrate only in an resonant excitation between two internal atomic levels. The atoms experience optical lattice potentials corresponding to the polarizability of each internal atomic state. The optical lattice potentials are in general different, but here we assume ground and excited state optical lattice potentials with minima at the same positions, which can be obtained for suitable choice of laser wave length \cite{Katori,Barber}. For a deep lattice, around the minimum of the optical lattice at each site the potential can be approximated by a harmonic potential with discrete levels \cite{Zoller}. Throughout the paper we assume the atoms to be localized at the ground state of these harmonic states, and we neglect excitations to higher levels. Note that already in previous works we studied collective electronic excitations (excitons) in such a system, and within a cavity we introduced polaritons \cite{ZoubiA,ZoubiB,ZoubiC,ZoubiD}.

The internal atomic transition between the ground state $|g\rangle$, of energy $\hbar\omega_g$, and the excited state $|e\rangle$, of energy $\hbar\omega_e$, has a transition energy of $\hbar\omega_A=\hbar\omega_e-\hbar\omega_g$. The expectation value of the transition dipole operator $\bar{\mu}$ is $\vec{\mu}=\langle e|\bar{\mu}|g\rangle$. The transition dipoles have a given orientation, with a direction fixed mainly in applying external static fields, e.g. electric or magnetic fields, and by the polarization of the optical lattice laser beams. The fact that the transition dipole has a fixed direction makes the system strongly anisotropic. A system of an optical lattice in the Mott insulator phase with one atom per site and a fixed transition dipole direction at each site, is similar to an artificial anisotropic crystal of Uniaxial type \cite{Born}. Identical considerations hold for molecular optical lattice of one molecule per site, e.g., for diatomic molecules the excited state has an oriented transition dipole, where due to the optical lattice polarization and in applying external static fields all the dipoles of different sites are organized in the same direction \cite{Rempe}.

In the present paper we study optical lattices with an anisotropy defined in the following, and through investigating optical spectra we achieve different physical properties of the system. To achieve our goals in an easily controllable system, the 2D anisotropic optical lattice is taken to be localized between two planar cavity mirrors \cite{ZoubiA}. For each cavity mode exists two possible orthogonal degenerate polarizations, TE and TM polarizations, of transverse electric and transverse magnetic fields, respectively \cite{Haroche}. We consider only a single perpendicular cavity mode, the one which is close to resonance to the above internal atomic transition. As the atomic transition dipole has a fixed direction, which is in general different from the direction of the cavity electric field, the coupling of the internal atomic excitations and the photons is a function of the angle between the transition dipoles and the photon polarizations. In the strong coupling regime the electronic excitations and the cavity photons of both polarizations are coherently mixed to form the new system diagonal eigenmodes, which are called cavity polaritons \cite{ZoubiE,Litinskaya}. Such photon polarization mixing can be observed via linear optical spectra \cite{ZoubiA,ZoubiE}. For an incident field which is, e.g., TE polarized, we expect, due to the anisotropic optical lattice, to observe transmitted and reflected fields of TE and TM polarizations. Photon polarization mixing can be also observed via the phase shift of the transmitted and reflected fields relative to an incident filed of a fixed polarization.

The paper is organized as follows. In section 1 we present anisotropic optical lattice within a cavity. The polarization mixing in the formation of cavity polaritons appears in section 2. Linear optical spectra is derived in section 3 with the optical light shift, for a given incident field. A summary is given in section 4.

\section{An anisotropic optical lattice within a cavity}

We consider a simple model of ultracold two-level atoms in an optical lattice in the Mott insulator phase with one atom per site. The atoms are prepared in a state, where the transition dipole moment matrix element has a fixed direction at each site, e.g., as in the case of figure (1). In fact we have multilevel atoms, but the previous case can be achieved in preparing all the ultracold atoms in a state of a fixed angular momentum, and to load them on an optical lattice with a fixed polarization accompanied with external static fields.

The atomic excitation Hamiltonian is given by
\begin{equation}
H_A=\sum_i\hbar\omega_A\ B_i^{\dagger}B_i,
\end{equation}
where $B_i^{\dagger}$ and $B_i$ are the creation and annihilation operators of electronic excitation at lattice site $i$, respectively. Here we neglect electrostatic interactions between atoms at different sites, as for small wave vectors they result only in an energy shift which can be absorbed in $\hbar\omega_A$, (their role in the formation of excitons was discussed widely by us in \cite{ZoubiA,ZoubiB}). In using the lattice symmetry, we can formally transform the excitation Hamiltonian into the momentum space in applying the transformation
\begin{equation}
B_i=\frac{1}{\sqrt{N}}\sum_{\bf q}e^{i{\bf q}\cdot{\bf n}_i}B_{\bf q},
\end{equation}
where $N=M\times M$ is the number of sites in the optical lattice, ${\bf n}_i$ is the position of site $i$, and ${\bf q}$ is the in-plane wave vector, which takes the values ${\bf q}=(q_x,q_y)=\frac{2\pi}{\sqrt{S}}(n_x,n_y)$, where $n_x,n_y=0,\pm 1,\pm 2,\cdots,\pm \frac{M}{2}$, with $S=Na^2$. The Hamiltonian is now written as
\begin{equation}
H_A=\sum_{\bf q}\hbar\omega_A\ B_{\bf q}^{\dagger}B_{\bf q},
\end{equation}

\begin{figure}
\centerline{\epsfxsize=6.0cm \epsfbox{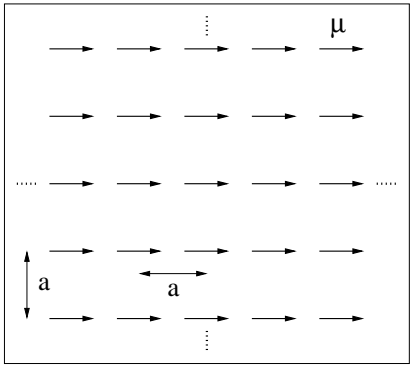}}
\caption{The optical lattice plane with oriented transition dipoles.}
\end{figure}

The 2D optical lattice is taken to be localized in the middle and parallel to planner cavity mirrors, as presented in figure (2). First we consider the case of perfect cavity mirrors, (later in order to get the linear optical spectra we consider the case of non-perfect mirrors). Here the electromagnetic field is free in the cavity plane with in-plane wave-vector ${\bf k}$, and is quantized in the perpendicular, $z$, direction with wave numbers $k_z=\frac{m\pi}{L}$, where $L$ is the distance between the cavity mirrors, and $m$ takes integer numbers, ($m=0,1,2,3,\cdots$). For each cavity mode $({\bf k}, m)$ we have two possible polarizations TE and TM, which are denoted by $(s)$ and $(p)$, respectively. The two cavity photon polarizations, $(\nu=s,p)$, are degenerate, where $\omega_{{\bf k}ms}=\omega_{{\bf k}mp}=\omega_{{\bf k}m}$, with the cavity photon dispersion
\begin{equation}
\omega_{{\bf k}m}=c\ \sqrt{k^2+\left(\frac{m\pi}{L}\right)^2},
\end{equation}
where $k=|{\bf k}|$. The cavity photon Hamiltonian is
\begin{equation}
H_C=\sum_{{\bf k},m,\nu}\hbar\omega_{{\bf k}m}\ a_{{\bf k}m\nu}^{\dagger}a_{{\bf k}m\nu},
\end{equation}
where $a_{{\bf k}m\nu}^{\dagger}$ and $a_{{\bf k}m\nu}$ are the creation and annihilation operators of a cavity photon in the mode $({{\bf k},m,\nu})$, respectively. 

The electric field operator is defined by
\begin{eqnarray}
{\bf E}({\bf r},z)&=&-i\sum_{{\bf k},m,\nu}\sqrt{\frac{\hbar\omega_{{\bf k}m}}{LS\epsilon_0}}\ \left\{{\bf u}^m_{\nu}({\bf k},z)e^{i{\bf k}\cdot{\bf r}}\ a_{{\bf k}m\nu}\right. \nonumber \\
&-&\left.{\bf u}^{m\ast}_{\nu}({\bf k},z)e^{-i{\bf k}\cdot{\bf r}}\ a_{{\bf k}m\nu}^{\dagger}\right\},
\end{eqnarray}
where $S$ is the cavity mirror quantization area, ${\bf r}$ is the in-plane position, and $z$ is the perpendicular position. The cavity mirrors are taken to be localized at the positions $z=\pm L/2$, see figure (2). The field vector functions are defined by \cite{Haroche}
\begin{equation}
{\bf u}^m_{s}({\bf k},z)=\sin\left[\frac{m\pi}{L}\left(z+\frac{L}{2}\right)\right]\ \hat{n}_{\bf k},
\end{equation}
where $m=0,1,2,\cdots$, and
\begin{eqnarray}
{\bf u}^m_{p}({\bf k},z)&=&-\frac{cm\pi}{L\omega_{{\bf k}m}}\left\{i\ \sin\left[\frac{m\pi}{L}\left(z+\frac{L}{2}\right)\right]\ \hat{e}_{\bf k}\right. \nonumber \\
&-&\left.\frac{kL}{m\pi}\ \cos\left[\frac{m\pi}{L}\left(z+\frac{L}{2}\right)\right]\ \hat{e}_{z}\right\},
\end{eqnarray}
where $m=1,2,3,\cdots$, and for $m=0$ we multiply ${\bf u}^0_{p}({\bf k},z)$ by the factor $1/\sqrt{2}$. The unit vectors are: $\hat{e}_{z}$ is along the $z$ axis, $\hat{e}_{\bf k}$ is along ${\bf k}$, that is $\hat{e}_{\bf k}={\bf k}/k$, and $\hat{n}_{\bf k}=\hat{e}_{\bf k}\times\hat{e}_{z}$, as illustrated in figure (3).

\begin{figure}
\centerline{\epsfxsize=6.0cm \epsfbox{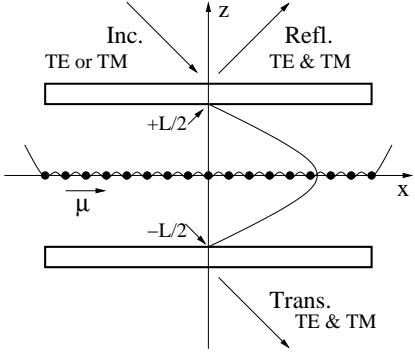}}
\caption{An optical lattice within a cavity.}
\end{figure}

\begin{figure}
\centerline{\epsfxsize=4.0cm \epsfbox{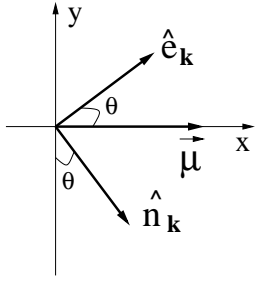}}
\caption{The transition dipole, and the unit vectors.}
\end{figure}

The optical lattice is located at the middle and parallel to the cavity mirrors at $z=0$. We assume only cavity modes with $m=1$ which are close to resonance to the atomic transition. We neglect all the other perpendicular modes. For multilevel atoms the cavity photons, of TE or TM linear polarizations, are close to resonance to the appropriate electronic transition with a fixed angular momentum. For the case of ground and/or excited states with degenerate multi-levels, where several allowed transitions close to resonance to the cavity photons of TE or TM polarizations, in applying external static fields the degenerate states split, and we can stay with a single electronic transition which is close to resonance to the cavity photons.

The coupling between the atomic transition and the cavity modes in the electric dipole approximation is given by the Hamiltonian $H_{AC}=-\bar{\mu}\cdot {\bf E}$, where the material electric dipole operator is defines as $\bar{\mu}=\sum_i\left(\vec{\mu}\ B_i^{\dagger}+\vec{\mu}^{\ast}\ B_i\right)$, and the electric field operator is evaluated at the atom positions. In the rotating wave approximation, the interaction is given by
\begin{eqnarray}
H_{AC}&=&i\sum_{{\bf k},\nu,i}\sqrt{\frac{\hbar\omega_{\bf k}}{LS\epsilon_0}}\left\{\left(\vec{\mu}\cdot{\bf u}_{\nu}({\bf k})\right)e^{i{\bf k}\cdot{\bf n}_i}\ B_i^{\dagger}a_{{\bf k}\nu}\right. \nonumber \\
&-&\left.\left(\vec{\mu}\cdot{\bf u}_{\nu}({\bf k})\right)^{\ast}e^{-i{\bf k}\cdot{\bf n}_i}\ a_{{\bf k}\nu}^{\dagger}B_i\right\}.
\end{eqnarray}
In transforming the electronic excitation operators into the momentum space we get
\begin{eqnarray}
H_{AC}&=&i\sum_{{\bf k},{\bf q},\nu,i}\sqrt{\frac{\hbar\omega_{\bf k}}{NLS\epsilon_0}}\left\{\left(\vec{\mu}\cdot{\bf u}_{\nu}({\bf k})\right)e^{i({\bf k-q})\cdot{\bf n}_i}\ B_{\bf q}^{\dagger}a_{{\bf k}\nu}\right. \nonumber \\
&-&\left.\left(\vec{\mu}\cdot{\bf u}_{\nu}({\bf k})\right)^{\ast}e^{-i({\bf k-q})\cdot{\bf n}_i}\ a_{{\bf k}\nu}^{\dagger}B_{\bf q}\right\}.
\end{eqnarray}
In using the property $\frac{1}{N}\sum_i e^{i({\bf k-q})\cdot{\bf n}_i}=\delta_{\bf k,q}$, we have
\begin{equation}
H_{AC}=\hbar \sum_{{\bf k},\nu}\left\{f_{\bf k}^{\nu}\ B_{\bf k}^{\dagger}a_{{\bf k}\nu}+f_{\bf k}^{\nu\ast}\ a_{{\bf k}\nu}^{\dagger}B_{\bf k}\right\},
\end{equation}
where the coupling parameter is
\begin{equation}
\hbar f_{\bf k}^{\nu}=i\sqrt{\frac{\hbar\omega_{\bf k}}{La^2\epsilon_0}}\ \left(\vec{\mu}\cdot{\bf u}_{\nu}({\bf k})\right).
\end{equation}
Due to the lattice translational symmetry, the coupling is between cavity photons and excitations with the same in-plane wave vector.

Explicitly, we have
\begin{eqnarray}
\hbar f_{\bf k}^{s}&=&i\sqrt{\frac{\hbar\omega_{\bf k}}{La^2\epsilon_0}}\ \left(\vec{\mu}\cdot\hat{n}_{\bf k}\right), \nonumber \\
\hbar f_{\bf k}^{p}&=&\sqrt{\frac{\hbar\omega_{\bf k}}{La^2\epsilon_0}}\left(\frac{\omega_0}{\omega_{\bf k}}\right)\ \left(\vec{\mu}\cdot\hat{e}_{\bf k}\right),
\end{eqnarray}
where $\omega_0=c\pi/L$. We interest in the case of small wave vectors ${\bf k}\approx {\bf 0}$, hence we neglected the contribution of the $z$ direction, (or we can assume $\mu_z\approx 0$). 

We take $\vec{\mu}$ to be real, e.g. for $\vec{\mu}=\mu\hat{x}$, (see figure (3)), and also we have
\begin{equation}
\hat{e}_{\bf k}=\cos\theta\ \hat{x}+\sin\theta\ \hat{y}\ ,\ \hat{n}_{\bf k}=\sin\theta\ \hat{x}-cos\theta\ \hat{y},
\end{equation}
to get
\begin{equation}
\hbar f_{\bf k}^{s}=iC_{\bf k}\ \sin\theta\ ,\ \hbar f_{\bf k}^{p}=C_{\bf k}\left(\frac{\omega_0}{\omega_{\bf k}}\right)\ \cos\theta,
\end{equation}
where $C_{\bf k}=\sqrt{\frac{\hbar\omega_{\bf k}\mu^2}{La^2\epsilon_0}}$. For example, if $\theta=\pi/4$, we get $\hbar f_{\bf k}^{s}=iC_{\bf k}/\sqrt{2}$, and $\hbar f_{\bf k}^{p}=C_{\bf k}\left(\frac{\omega_0}{\omega_{\bf k}\sqrt{2}}\right)$. For $\theta=0$, we get $\hbar f_{\bf k}^{s}=0$, and $\hbar f_{\bf k}^{p}=C_{\bf k}\left(\frac{\omega_0}{\omega_{\bf k}}\right)$. For $\theta=\pi/2$, we get $\hbar f_{\bf k}^{s}=iC_{\bf k}$, and $\hbar f_{\bf k}^{p}=0$.

In the case of cavity photons of standing waves without propagations, we need only to substitute ${\bf k}=0$ in the above results. The cavity modes are of transverse electric and magnetic fields, that is TEM modes, with two orthogonal polarizations, which are denoted here also by $(s)$ and $(p)$. For $(m=1)$ the coupling parameters are $\hbar f_0^{s}=iC_0\sin\theta$ and $\hbar f_0^{p}=C_0\cos\theta$, where $C_0=\sqrt{\frac{\hbar\omega_0\mu^2}{La^2\epsilon_0}}$.

\section{Polarization mixing in the strong coupling regime}

The total Hamiltonian of the coupled electronic excitations and cavity photons is given by
\begin{eqnarray}
H&=&\hbar\sum_{\bf k}\left\{\omega_A\ B_{\bf k}^{\dagger}B_{\bf k}+\sum_{\nu}\omega_{\bf k}\ a_{{\bf k}\nu}^{\dagger}a_{{\bf k}\nu}\right. \nonumber \\
&+&\left.\sum_{\nu}\left(f_{\bf k}^{\nu}\ B_{\bf k}^{\dagger}a_{{\bf k}\nu}+f_{\bf k}^{\nu\ast}\ a_{{\bf k}\nu}^{\dagger}B_{\bf k}\right)\right\}.
\end{eqnarray}
In the strong coupling regime, where the coupling is larger than the atomic excitation and the cavity photon line-width, the real system eigenmodes are cavity polaritons \cite{ZoubiA,ZoubiE}, which are obtained in diagonalizing the above Hamiltonian, to get
\begin{equation}
H=\sum_{{\bf k},r}\hbar\Omega_{{\bf k}r}\ A_{{\bf k}r}^{\dagger}A_{{\bf k}r},
\end{equation}
where we obtain three polariton branches, with the eigenfrequencies
\begin{equation}
\Omega_{{\bf k}\pm}=\frac{\omega_{\bf k}+\omega_A}{2}\pm\Delta_{\bf k}\ ,\ \Omega_{{\bf k}0}=\omega_{\bf k},
\end{equation}
where 
\begin{equation}
\Delta_{\bf k}=\sqrt{\delta_{\bf k}^2+|f_{\bf k}|^2}\ , \ |f_{\bf k}|^2=\sum_{\nu}|f_{\bf k}^{\nu}|^2,
\end{equation}
with the excitation-photon detuning
\begin{equation}
\delta_{\bf k}=\frac{\omega_{\bf k}-\omega_A}{2}.
\end{equation}

For the case of atomic transition frequency of $\omega_A/2\pi=2.5\times10^{14}\ Hz$, and for the first cavity mode $m=1$ with a distance between the cavity mirrors of $L=c\pi/\omega_0\approx 3.77\ \mu m$, where we get zero detuning between the atomic transition and the cavity photon at $k=0$, we plot in figure (4) the three polariton frequency dispersions, $\Omega_{{\bf k}r}/2\pi$, as a function of $k$, at the angle $\theta=\pi/4$, and for transition dipole of $\mu=2\ e\AA$, and lattice constant of $a=2000\ \AA=0.2\ \mu m$. 

\begin{figure}
\centerline{\epsfxsize=7.0cm \epsfbox{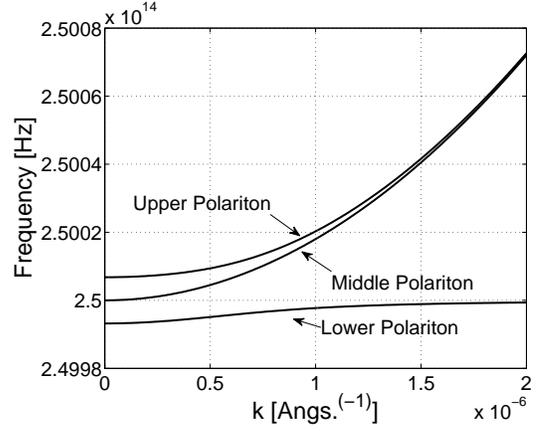}}
\caption{The three polariton frequency dispersions $\Omega_{r}/2\pi$ vs. $k$, at $\theta=\pi/4$.}
\end{figure}

In the limit of large detuning, where $\delta_{\bf k}\gg|f_{\bf k}|$, we get $\Omega_{{\bf k}+}\approx\omega_{\bf k}+\frac{|f_{\bf k}|^2}{2\delta_{\bf k}}$, $\Omega_{{\bf k}-}\approx\omega_A-\frac{|f_{\bf k}|^2}{2\delta_{\bf k}}$, and $\Omega_{{\bf k}0}=\omega_{\bf k}$. The upper branch is a photon with a shift due to the coupling to the excitation, and the lower is an excitation with a light shift due to the coupling to the cavity photon. Here the middle branch still a photon as before. In this limit we get birefringence, as we have two refracted cavity fields, the ordinary field of the $(0)$ branch, and the extraordinary field of the $(+)$ branch.

The polariton operators are in general a coherent superposition of atomic excitations and cavity photons of both polarizations, namely we have
\begin{equation}
A_{{\bf k}r}=X_{{\bf k}r}\ B_{\bf k}+\sum_{\nu}Y_{{\bf k}r}^{\nu}\ a_{{\bf k}\nu},
\end{equation}
with the relation $|X_{{\bf k}r}|^2+\sum_{\nu}|Y_{{\bf k}r}^{\nu}|^2=1$, where the excitation and photon amplitudes of the upper $(+)$ and the lower $(-)$ polariton branches are
\begin{equation}
X_{{\bf k}\pm}=\pm\sqrt{\frac{\Delta_{\bf k}\mp\delta_{\bf k}}{2\Delta_{\bf k}}}\ ,\ Y_{{\bf k}\pm}^{\nu}=\frac{f_{\bf k}^{\nu}}{\sqrt{2\Delta_{\bf k}(\Delta_{\bf k}\mp\delta_{\bf k})}},
\end{equation}
and the excitation and photon amplitudes of the middle $(0)$ polariton branch are
\begin{equation}
X_{{\bf k}0}=0\ ,\ Y_{{\bf k}0}^{\nu}=\frac{f_{\bf k}^{\nu}}{|f_{\bf k}|}.
\end{equation}
Note that the middle polariton branch is pure photonic. Exist a direction, for each ${\bf k}$, where the polarization direction of a photon, which is a superposition of cavity photons of both polarizations, is orthogonal to the transition dipole, and hence no interaction between this photon and the material is obtained. In other words, as seen in figure (3), the superposition of the two photon polarizations gives two components. The longitudinal component along the transition dipole, which forms in the strong coupling regime with the excitation the two upper and lower polariton branches. The transverse component which is orthogonal to the transition dipole and decouple to the excitation, to form the photonic middle branch.

At the intersection point between the transition frequency and the cavity photon dispersion, where $\delta_{\bf k}=0$, the polaritons are half excitation and half photon, namely $|X_{{\bf k}r}|^2=1/2$ and $\sum_{\nu}|Y_{{\bf k}r}^{\nu}|^2=1/2$. At large wave vectors the upper branch became photonic, with $|X_{{\bf k}r}|^2\approx 0$ and $\sum_{\nu}|Y_{{\bf k}r}^{\nu}|^2\approx 1$, and the lower branch became excitation, with $|X_{{\bf k}r}|^2\approx 1$ and $\sum_{\nu}|Y_{{\bf k}r}^{\nu}|^2\approx 0$.

In figures (5-9) we plot the excitation and photon weights in the three polariton branches, $|X_{{\bf k}r}|^2$ and $|Y_{{\bf k}r}^{\nu}|^2$, as a function of $k$, for the angle $\theta=\pi/4$. It is seen that for small $k$ the upper and lower polariton branches, in figures (5,8), are a coherent mix of the excitation and the cavity photon of both polarizations. The upper branch, for large $k$, becomes photonic, and for larger $k$ it becomes photon of $(s)$ polarization, as seen in figure (5) and more clearly in figures (6-7). While the lower branch becomes excitation for large $k$, as appears in figure (8). The middle branch is pure photonic, as seen in figure (9), and for large $k$ becomes $(s)$ polarized photon. In figures (10-12) we plot the excitation and photon weights as a function of the angle $\theta$ at $k=5\times10^{-7}\ \AA^{-1}$. The plots indicate the significant role of the angle controllability.

\begin{figure}
\centerline{\epsfxsize=7.0cm \epsfbox{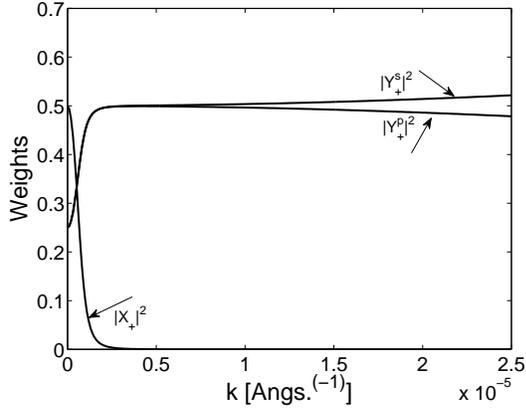}}
\caption{The excitation and photon of both polarization weights in the upper branch, $|X_{{\bf k}+}|^2$, $|Y_{{\bf k}+}^s|^2$, and $|Y_{{\bf k}+}^p|^2$ vs. $k$, at $\theta=\pi/4$.}
\end{figure}

\begin{figure}
\centerline{\epsfxsize=7.0cm \epsfbox{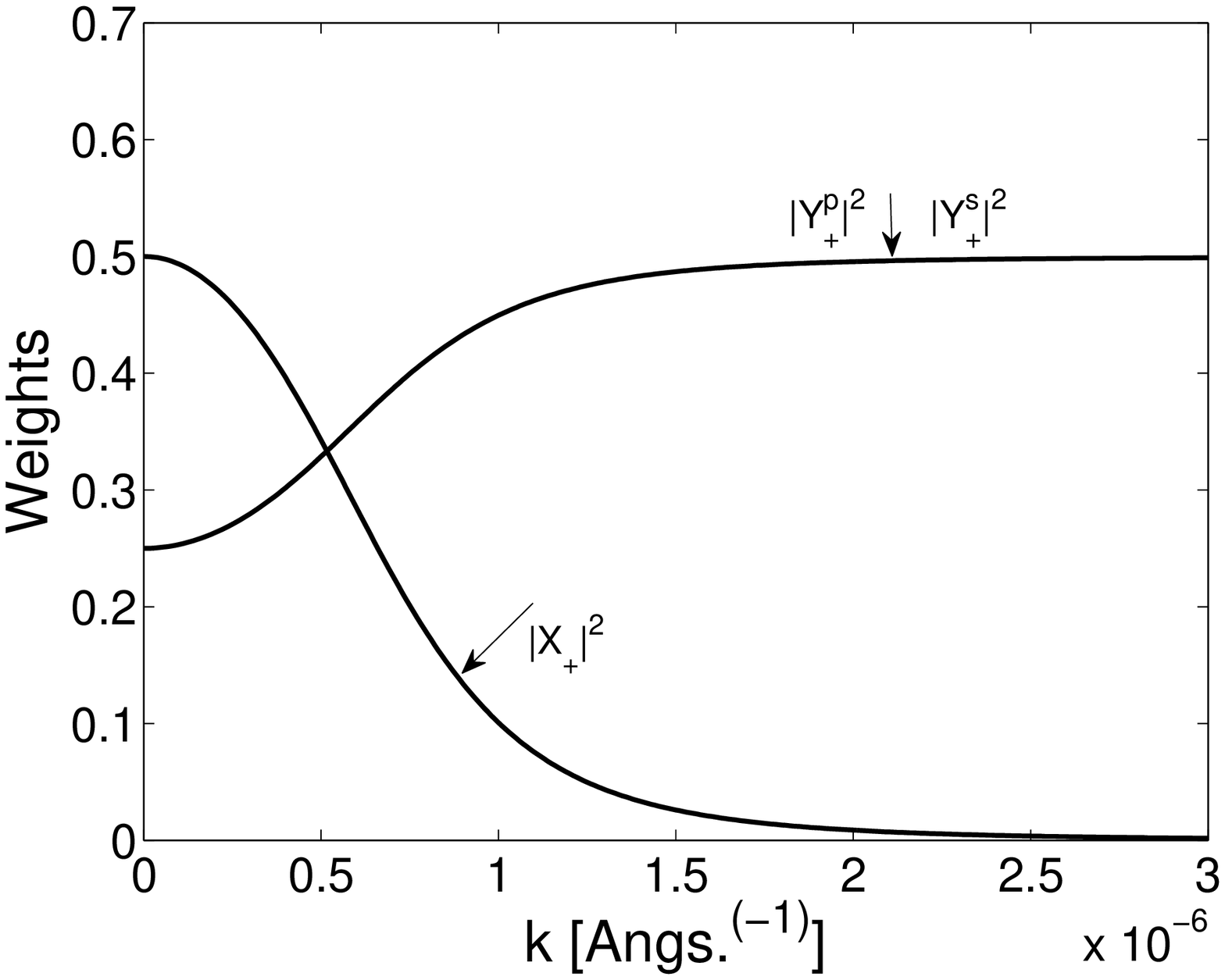}}
\caption{The excitation and photon of both polarization weights in the upper branch, $|X_{{\bf k}+}|^2$, $|Y_{{\bf k}}^s|^2$, and $|Y_{{\bf k}+}^p|^2$ vs. $k$, for small $k$ and at $\theta=\pi/4$.}
\end{figure}

\begin{figure}
\centerline{\epsfxsize=7.0cm \epsfbox{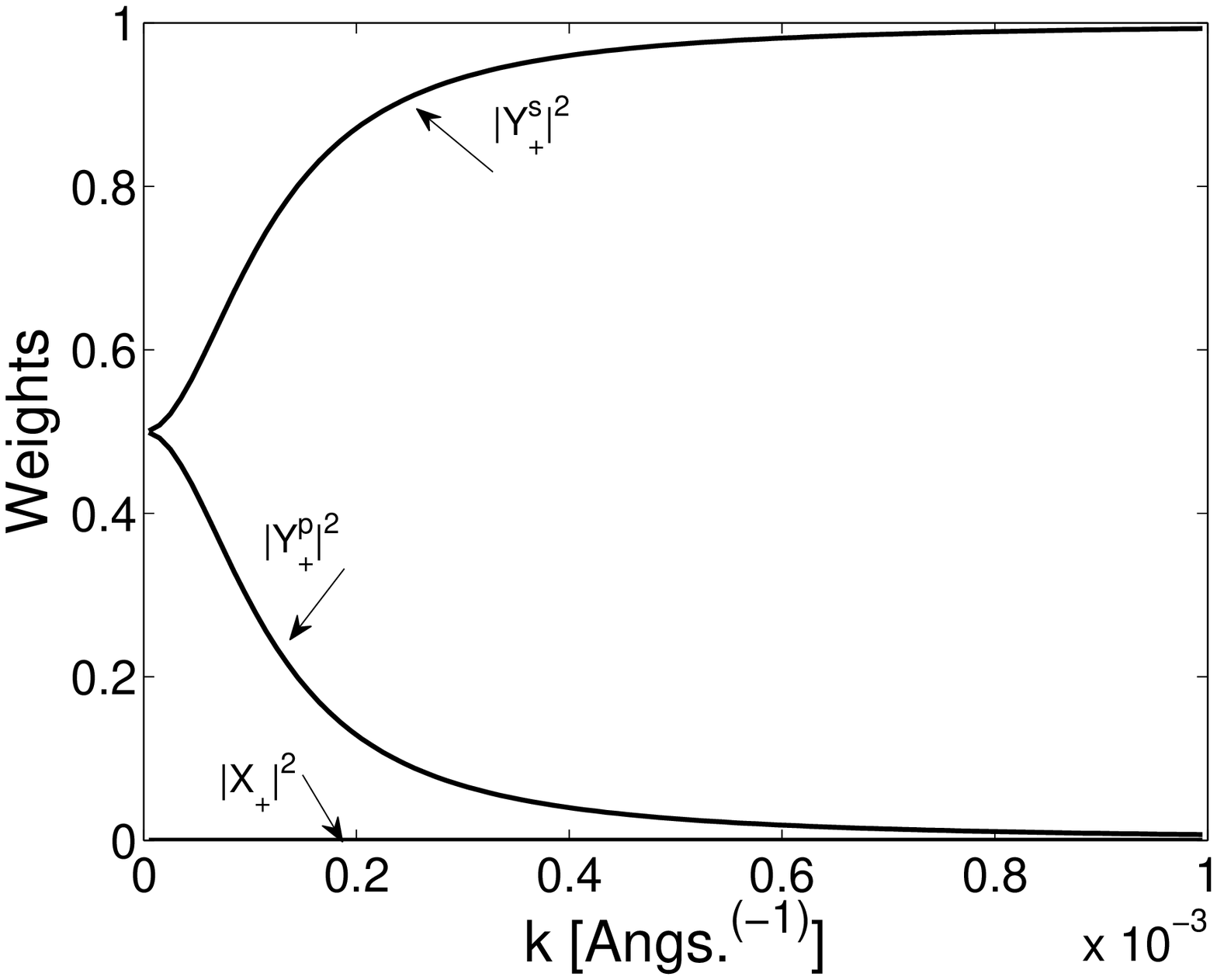}}
\caption{The excitation and photon of both polarization weights in the upper branch, $|X_{{\bf k}+}|^2$, $|Y_{{\bf k}+}^s|^2$, and $|Y_{{\bf k}+}^p|^2$ vs. $k$, for larger $k$ and at $\theta=\pi/4$.}
\end{figure}

\begin{figure}
\centerline{\epsfxsize=7.0cm \epsfbox{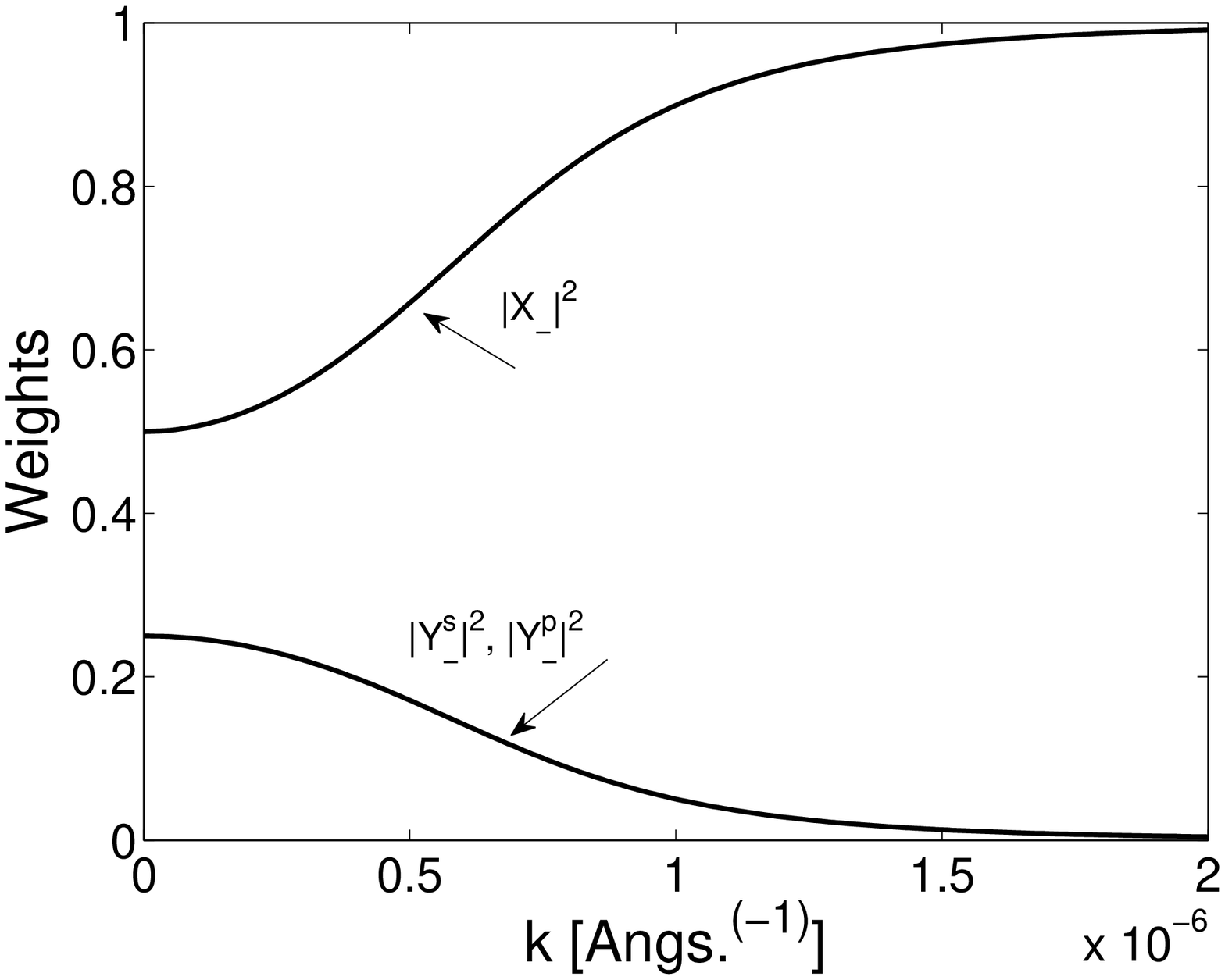}}
\caption{The excitation and photon of both polarization weights in the lower branch, $|X_{{\bf k}-}|^2$, $|Y_{{\bf k}-}^s|^2$, and $|Y_{{\bf k}-}^p|^2$ vs. $k$, at $\theta=\pi/4$.}
\end{figure}

\begin{figure}
\centerline{\epsfxsize=7.0cm \epsfbox{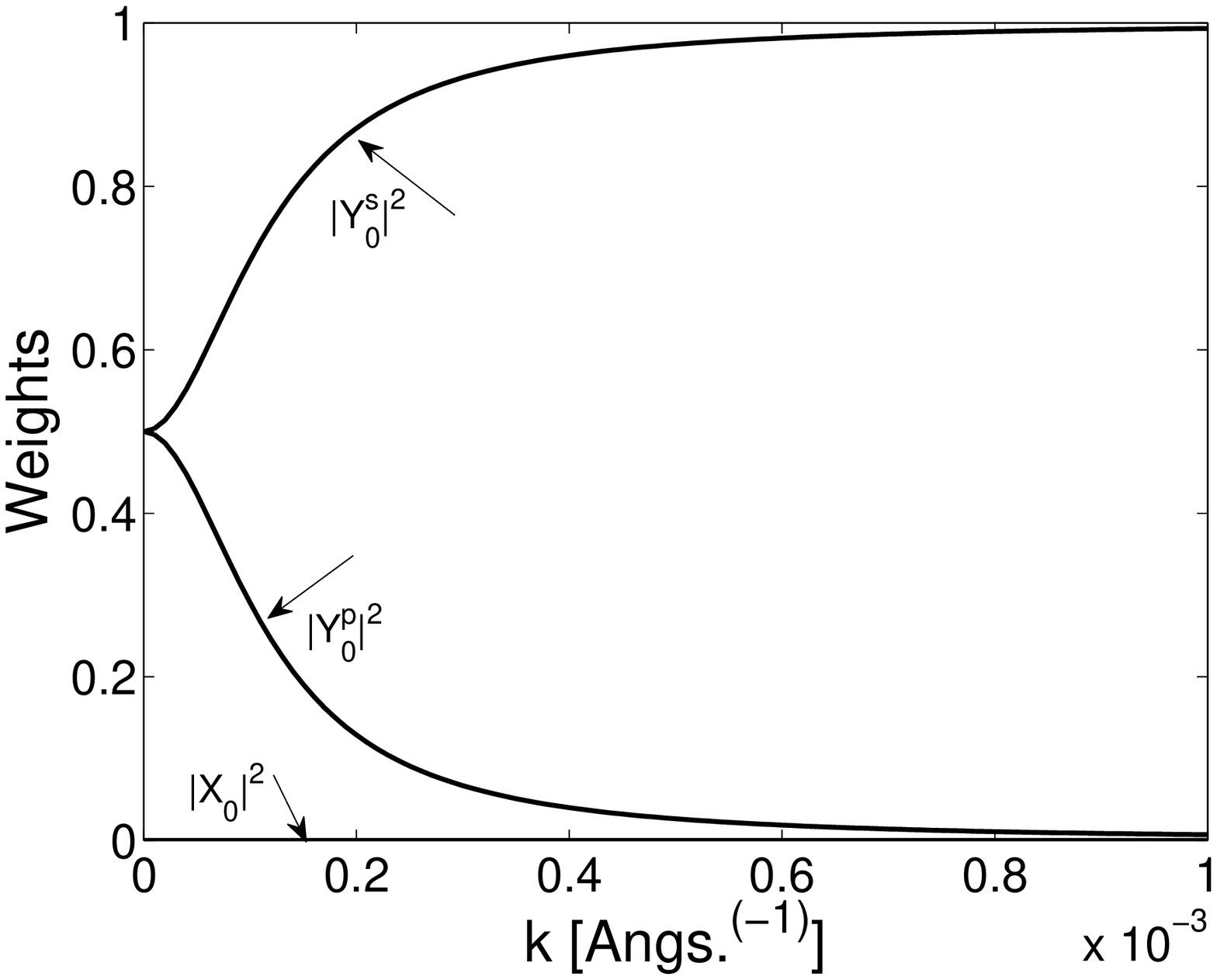}}
\caption{The excitation and photon of both polarization weights in the middle branch, $|X_{{\bf k}0}|^2$, $|Y_{{\bf k}0}^s|^2$, and $|Y_{{\bf k}0}^p|^2$ vs. $k$, at $\theta=\pi/4$.}
\end{figure}

\begin{figure}
\centerline{\epsfxsize=7.0cm \epsfbox{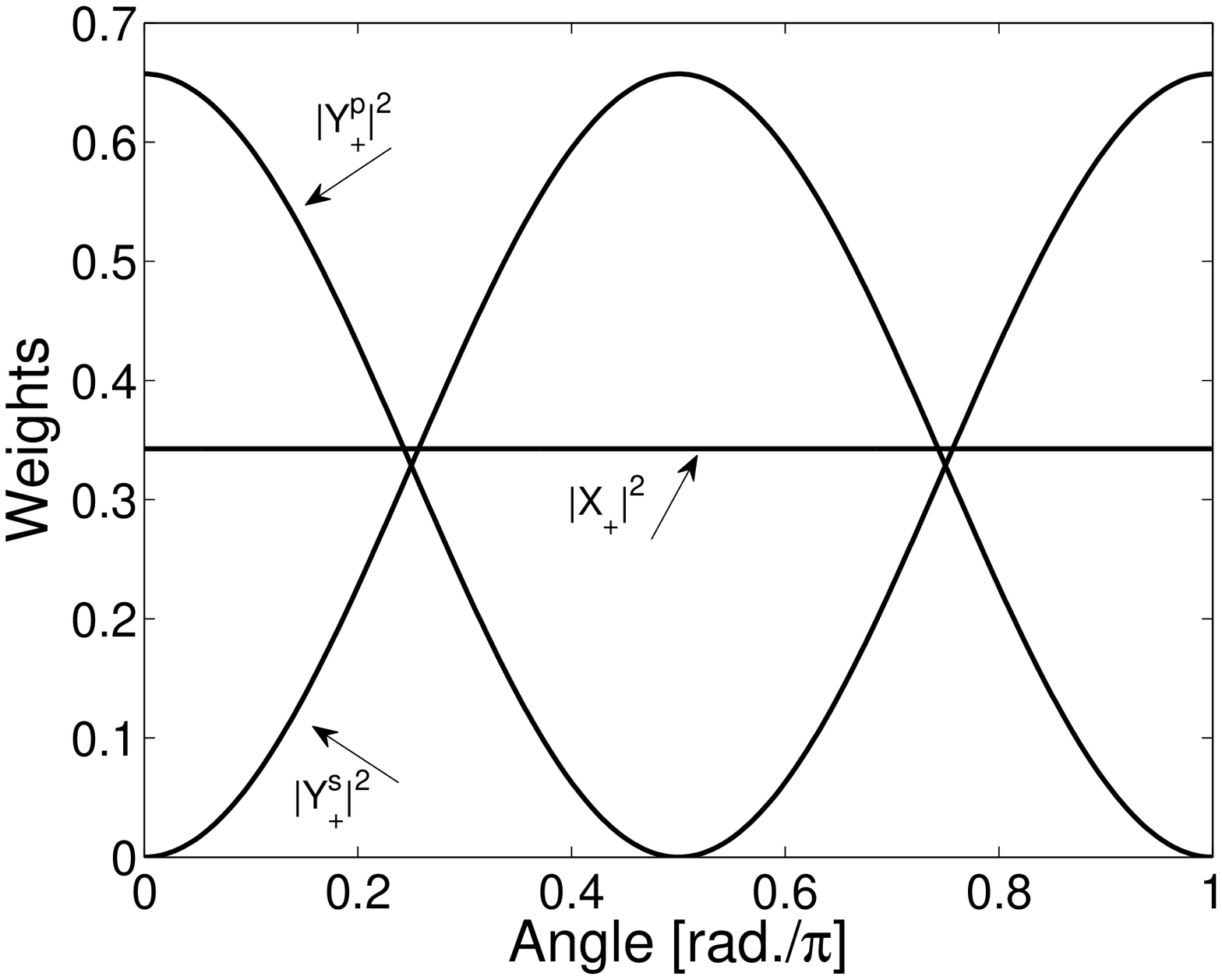}}
\caption{The excitation and photon of both polarization weights in the upper branch, $|X_{{\bf k}+}|^2$, $|Y_{{\bf k}+}^s|^2$, and $|Y_{{\bf k}+}^p|^2$ vs. $\theta$, at $k=5\times10^{-7}\ \AA^{-1}$.}
\end{figure}

\begin{figure}
\centerline{\epsfxsize=7.0cm \epsfbox{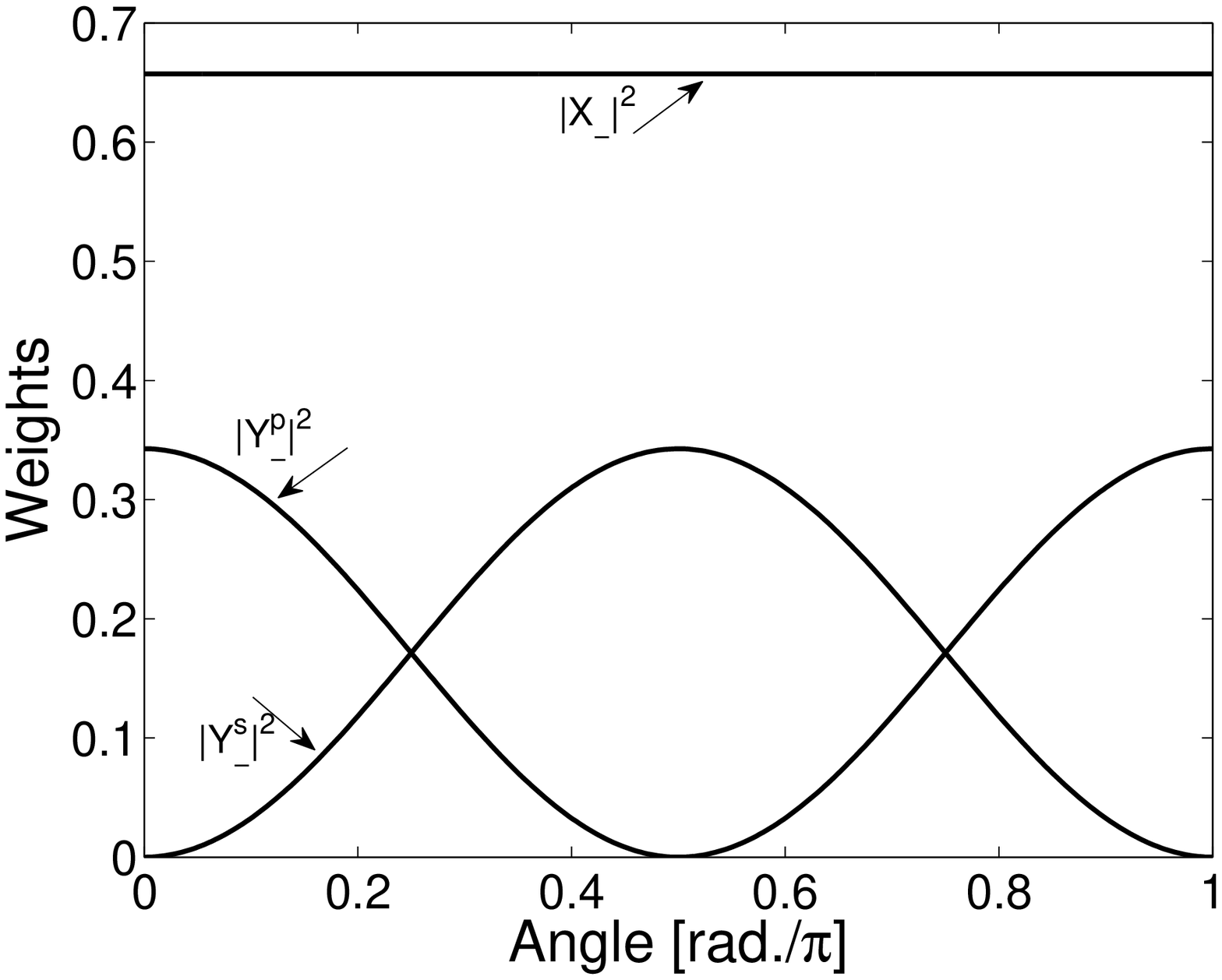}}
\caption{The excitation and photon of both polarization weights in the lower branch, $|X_{{\bf k}-}|^2$, $|Y_{{\bf k}-}^s|^2$, and $|Y_{{\bf k}-}^p|^2$ vs. $\theta$, at $k=5\times10^{-7}\ \AA^{-1}$.}
\end{figure}

\begin{figure}
\centerline{\epsfxsize=7.0cm \epsfbox{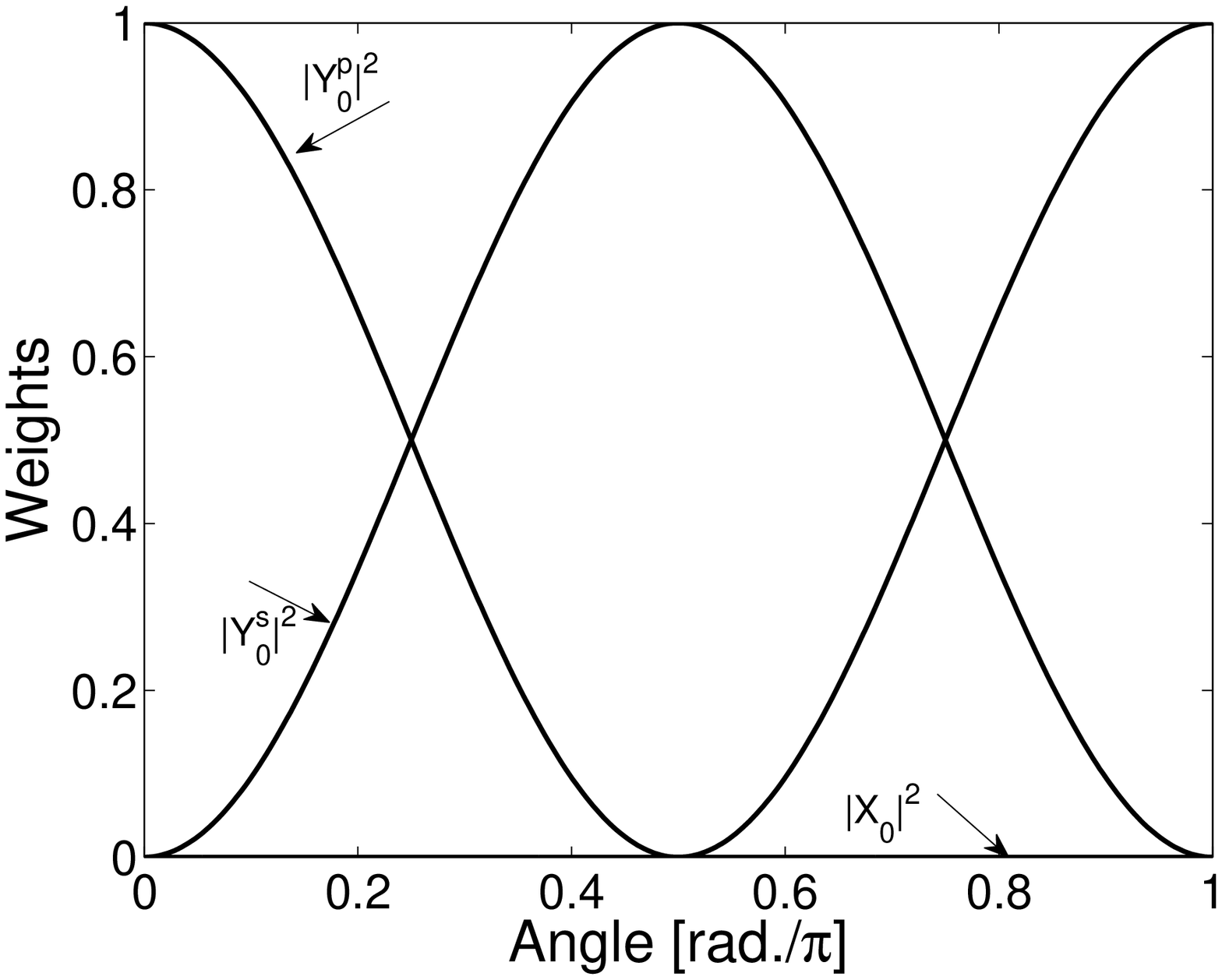}}
\caption{The excitation and photon of both polarization weights in the middle branch, $|X_{{\bf k}0}|^2$, $|Y_{{\bf k}0}^s|^2$, and $|Y_{{\bf k}0}^p|^2$ vs. $\theta$, at $k=5\times10^{-7}\ \AA^{-1}$.}
\end{figure}

\section{Anisotropic Linear Optical Spectra}

To observe the system eigenmodes we need to couple the internal system to the external world. Our observation tool is Linear Optical Spectra \cite{ZoubiA,ZoubiE}. In assuming non-perfect mirrors, the internal system get coupled to their environment. For an incident external field with a fixed in-plane wave vector and polarization we calculate the Transmission, Reflection, and Absorption Spectra, as seen in figure (2). Now the electromagnetic field is divided into two parts, the cavity field and the environment field, and they coupled through the non-perfect cavity mirrors. The life time of the excited state is included phenomenologically.

The external electromagnetic field is given by the following Hamiltonians, for fields at the two sides of the cavity, 
\begin{equation}
H_U=\sum_{{\bf k},\nu}\int d\omega_{\bf k}\ \hbar\omega_{\bf k}\ b_{{\bf k}\nu}^{\dagger}(\omega_{\bf k})b_{{\bf k}\nu}(\omega_{\bf k}),
\end{equation}
and
\begin{equation}
H_L=\sum_{{\bf k},\nu}\int d\omega_{\bf k}\ \hbar\omega_{\bf k}\ c_{{\bf k}\nu}^{\dagger}(\omega_{\bf k})c_{{\bf k}\nu}(\omega_{\bf k}),
\end{equation}
where $b_{{\bf k}\nu}^{\dagger}(\omega_{\bf k})$ and $b_{{\bf k}\nu}(\omega_{\bf k})$ are the creation and annihilation operators of an external photon in the upper side of the cavity, and $c_{{\bf k}\nu}^{\dagger}(\omega_{\bf k})$ and $c_{{\bf k}\nu}(\omega_{\bf k})$ are the creation and annihilation operators of an external photon in the lower side of the cavity. The external photon energy is $\hbar\omega_{\bf k}$. The coupling of the cavity photons to the external fields is given by
\begin{equation}
V_U=\sum_{{\bf k},\nu}\int d\omega_{\bf k}\ i\hbar u(\omega_{\bf k})\ \left\{b_{{\bf k}\nu}^{\dagger}(\omega_{\bf k})a_{{\bf k}\nu}-a_{{\bf k}\nu}^{\dagger}b_{{\bf k}\nu}(\omega_{\bf k})\right\},
\end{equation}
and 
\begin{equation}
V_L=\sum_{{\bf k},\nu}\int d\omega_{\bf k}\ i\hbar v(\omega_{\bf k})\ \left\{c_{{\bf k}\nu}^{\dagger}(\omega_{\bf k})a_{{\bf k}\nu}-a_{{\bf k}\nu}^{\dagger}c_{{\bf k}\nu}(\omega_{\bf k})\right\},
\end{equation}
where $u(\omega_{\bf k})$ is the coupling parameter through the upper mirror, and $v(\omega_{\bf k})$ through the lower mirror. The coupling is between cavity photons and external photons with the same in-plane wave vector and polarization. No polarization mixing or wave vector scattering can be obtained due to the cavity mirrors.

The total Hamiltonian of the coupled internal system and external field is separable for each in-plane wave vector ${\bf k}$, hence we can treat the whole system for each ${\bf k}$ separately. The Hamiltonian for a fixed ${\bf k}$, where we can drop ${\bf k}$ in the following, is given by
\begin{eqnarray}
H&=&\sum_{r}\hbar\Omega_{r}\ A_{r}^{\dagger}A_{r}+\sum_{\nu}\int d\omega\ \hbar\omega\ b_{\nu}^{\dagger}(\omega)b_{\nu}(\omega) \nonumber \\
&+&\sum_{\nu}\int d\omega\ \hbar\omega\ c_{\nu}^{\dagger}(\omega)c_{\nu}(\omega) \nonumber \\
&+&\sum_{r,\nu}\int d\omega\ i\hbar u(\omega)\ \left\{Y_{\nu}^{r\ast}\ b_{\nu}^{\dagger}(\omega)A_{r}-Y_{\nu}^{r}\ A_{r}^{\dagger}b_{\nu}(\omega)\right\} \nonumber \\
&+&\sum_{r,\nu}\int d\omega\ i\hbar v(\omega)\ \left\{Y_{\nu}^{r\ast}\ c_{\nu}^{\dagger}(\omega)A_{r}-Y_{\nu}^{r}\ A_{r}^{\dagger}c_{\nu}(\omega)\right\}, \nonumber \\
\end{eqnarray}
where we used the cavity photon operator in terms of polariton operators, in using the inverse transformation $a_{\nu}=\sum_r Y_{\nu}^{r\ast}\ A_{r}$.

The equations of motion for the external field operators are
\begin{eqnarray}
\frac{d}{dt}b_{\nu}(\omega)&=&-i\omega\ b_{\nu}(\omega)+u(\omega)\ \sum_rY_{\nu}^{r\ast}\ A_r, \nonumber \\
\frac{d}{dt}c_{\nu}(\omega)&=&-i\omega\ c_{\nu}(\omega)+v(\omega)\ \sum_rY_{\nu}^{r\ast}\ A_r,
\end{eqnarray}
and for the polariton operator we get
\begin{eqnarray}
&&\frac{d}{dt}A_r=-i\Omega_{r}\ A_r \nonumber \\
&-&\sum_{\nu}\int d\omega\ Y_{\nu}^{r}\ \left\{u(\omega)\ b_{\nu}(\omega)+v(\omega)\ c_{\nu}(\omega)\right\}.
\end{eqnarray}

Using the input-output formalisms \cite{GardinerA}, the equations for the external field operators are solved formally for the initial and final conditions and substituted back in the polariton equation, and after applying the Markov approximation \cite{GardinerB}, we get the two equations
\begin{eqnarray}
\frac{d}{dt}A_r&=&-i\Omega_{r}\ A_r-\gamma\sum_{\nu}\ Y_{\nu}^{r}\ a_{\nu} \nonumber \\
&+&\sum_{\nu}\ Y_{\nu}^{r}\ \left\{\sqrt{\gamma_U}\ b^{\nu}_{in}+\sqrt{\gamma_L}\ c^{\nu}_{in}\right\},
\end{eqnarray}
and
\begin{eqnarray}
\frac{d}{dt}A_r&=&-i\Omega_{r}\ A_r+\gamma\sum_{\nu}\ Y_{\nu}^{r}\ a_{\nu} \nonumber \\
&-&\sum_{\nu}\ Y_{\nu}^{r}\ \left\{\sqrt{\gamma_U}\ b^{\nu}_{out}+\sqrt{\gamma_L}\ c^{\nu}_{out}\right\},
\end{eqnarray}
where we defined the damping rates at the two mirrors to be constants, and are given by $\gamma_U=2\pi\ u^2(\omega)$, and $\gamma_L=2\pi\ v^2(\omega)$, and we defined also $\gamma=\frac{\gamma_U+\gamma_L}{2}$. Furthermore, we defined the input and output fields at the upper mirror by $b^{\nu}_{in}$ and $b^{\nu}_{out}$, and at the lower mirror by $c^{\nu}_{in}$ and $c^{\nu}_{out}$, respectively.

The boundary condition between the cavity and the external fields at the upper and lower mirrors are given by
\begin{equation}
\sqrt{\gamma_U}\ a_{\nu}=b^{\nu}_{in}+b^{\nu}_{out}\ ,\ \sqrt{\gamma_L}\ a_{\nu}=c^{\nu}_{in}+c^{\nu}_{out}.
\end{equation}

The electronic excitation damping rate is $\Gamma_{ex}$. Phenomenologically, the $r$ polariton branch damping rate is $\Gamma_r=\Gamma_{ex}\ |X^r|^2$, where $|X^r|^2$ is the excitation weight of the $(r)$ polariton. We define the polariton complex frequency by $\bar{\Omega}_r=\Omega_r-i\Gamma_r$.

In applying the Fourier transform, from time $t$ into frequency $\omega$ space, we get the system of equations
\begin{eqnarray}
i(\bar{\Omega}_r-\omega)\ \tilde{A}_r&=&\gamma\sum_{\nu}\ Y_{\nu}^{r}\ \tilde{a}_{\nu} \nonumber \\
&-&\sum_{\nu}\ Y_{\nu}^{r}\ \left\{\sqrt{\gamma_U}\ \tilde{b}^{\nu}_{out}+\sqrt{\gamma_L}\ \tilde{c}^{\nu}_{out}\right\}, \nonumber \\
i(\bar{\Omega}_r-\omega)\ \tilde{A}_r&=&-\gamma\sum_{\nu}\ Y_{\nu}^{r}\ \tilde{a}_{\nu} \nonumber \\
&+&\sum_{\nu}\ Y_{\nu}^{r}\ \left\{\sqrt{\gamma_U}\ \tilde{b}^{\nu}_{in}+\sqrt{\gamma_L}\ \tilde{c}^{\nu}_{in}\right\},
\end{eqnarray}
and
\begin{equation}
\sqrt{\gamma_U}\ \tilde{a}_{\nu}=\tilde{b}^{\nu}_{in}+\tilde{b}^{\nu}_{out}\ ,\ \sqrt{\gamma_L}\ \tilde{a}_{\nu}=\tilde{c}^{\nu}_{in}+\tilde{c}^{\nu}_{out}.
\end{equation}

Two assumptions will be done here, in order to simplify the system of equations. The first is to assume an input external field from only the upper mirror, namely we have $\tilde{c}^{\nu}_{in}=0$. Second, we assume the upper and lower mirrors to be identical, namely we have $\gamma_U=\gamma_L=\gamma$. Hence, in term of cavity and external photon operators, we get the system of equations
\begin{eqnarray}
\tilde{a}_{\alpha}&=&\sum_{\beta}\Lambda_{\alpha\beta}\ \left\{\gamma\ \tilde{a}_{\beta}-\sqrt{\gamma}\ \left(\tilde{b}^{\beta}_{out}-\tilde{c}^{\beta}_{out}\right)\right\}, \nonumber \\
\tilde{a}_{\alpha}&=&\sum_{\beta}\Lambda_{\alpha\beta}\ \left\{-\gamma\ \tilde{a}_{\beta}+\sqrt{\gamma}\ \tilde{b}^{\beta}_{in}\right\},
\end{eqnarray}
and
\begin{equation}
\sqrt{\gamma}\ \tilde{a}_{\alpha}=\tilde{b}^{\alpha}_{in}+\tilde{b}^{\alpha}_{out}\ ,\ \sqrt{\gamma}\ \tilde{a}_{\alpha}=\tilde{c}^{\alpha}_{out},
\end{equation}
where we defined the matrix
\begin{equation}
\Lambda_{\alpha\beta}=i\sum_r\frac{Y_{\alpha}^{r\ast}Y_{\beta}^{r}}{\omega-\bar{\Omega}_r}.
\end{equation}
Here we will solve the above system of equation for the following case.

For incident field of only $(s)$ polarization, where $\tilde{b}^{p}_{in}=0$, the solution is
\begin{eqnarray} \label{SPOL}
\frac{\tilde{c}^{s}_{out}}{\tilde{b}^{s}_{in}}&=&\frac{\gamma\ \Lambda_{ss}(1+\gamma\ \Lambda_{pp})-\gamma^2\ \Lambda_{sp}\Lambda_{ps}}{D}=t^{(s)}_s\ e^{i\Delta\phi^{(s)t}_s}, \nonumber \\
\frac{\tilde{c}^{p}_{out}}{\tilde{b}^{s}_{in}}&=&\frac{\tilde{b}^{p}_{out}}{\tilde{b}^{s}_{in}}=\frac{\gamma\ \Lambda_{ps}}{D}=t^{(s)}_p\ e^{i\Delta\phi^{(s)t}_p}=r^{(s)}_p\ e^{i\Delta\phi^{(s)r}_p}, \nonumber \\
\frac{\tilde{b}^{s}_{out}}{\tilde{b}^{s}_{in}}&=&\frac{-(1+\gamma\ \Lambda_{pp})}{D}=r^{(s)}_s\ e^{i\Delta\phi^{(s)r}_s},
\end{eqnarray}
where
\begin{equation}
D=(1+\gamma\ \Lambda_{ss})(1+\gamma\ \Lambda_{pp})-\gamma^2\ \Lambda_{sp}\Lambda_{ps}.
\end{equation}
The transmission and reflection of $(s)$ polarized fields are
\begin{eqnarray}
T_s^{(s)}&=&\left(t^{(s)}_s\right)^2=\frac{\gamma^2\ |\Lambda_{ss}(1+\gamma\ \Lambda_{pp})-\gamma\ \Lambda_{sp}\Lambda_{ps}|^2}{|D|^2}, \nonumber \\
R_s^{(s)}&=&\left(r^{(s)}_s\right)^2=\frac{|1+\gamma\ \Lambda_{pp}|^2}{|D|^2}.
\end{eqnarray}
The transmission and reflection of $(p)$ polarized fields are
\begin{equation}
T_p^{(s)}=R_p^{(s)}=\left(t^{(s)}_p\right)^2=\left(r^{(s)}_p\right)^2=\frac{\gamma^2\ |\Lambda_{ps}|^2}{|D|^2}.
\end{equation}
Even though the incident field is $(s)$ polarized, due to the anisotropy in the optical lattice we get transmitted and reflected fields which are $(p)$ polarized. Moreover, due to the assumption of identical mirrors, the $(p)$ polarized transmission and reflection fields are equal.

The absorption, $A$, in the cavity medium is calculated from the identity relation
\begin{equation}
T_s^{(s)}+T_p^{(s)}+R_s^{(s)}+R_p^{(s)}+A^{(s)}=1.
\end{equation}

Also from the system of equations (\ref{SPOL}) we deduce the phase shift in the $(s)$ and $(p)$ polarized transmitted fields, $\Delta\phi^{(s)t}_s$ and $\Delta\phi^{(s)t}_p$, and reflected fields, $\Delta\phi^{(s)r}_s$ and $\Delta\phi^{(s)r}_p$, relative to the incident field, respectively.

In figures (13-14) we plot the transmission and reflection spectra of the $(s)$ polarized fields, $T_s^{(s)}$ and $R_s^{(s)}$, as a function of frequency, $\omega\rightarrow\omega/2\pi$, respectively, and for different angles $\theta$, at $k=5\times10^{-7}\ \AA^{-1}$. The three peaks and three dips correspond to the three polariton branches. At $\theta=0$ we get zero transmission and complete reflection. The largest transmission peaks and the deepest reflection dips are obtained at $\theta=\pi/2$. In figure (15) we plot the transmission and reflection spectra of the $(p)$ polarized fields, $T_p^{(s)}$ and $R_p^{(s)}$, which are equal for identical cavity mirrors. Even though the incident field is $(s)$ polarized we get transmission and reflection of $(p)$ polarized fields, with maximum at $\theta=\pi/4$, and zeros at $\theta=0$ and $\theta=\pi/2$. In figure (16) we plot the absorption spectra $A^{(s)}$. Only two peaks are obtained correspond to the upper and lower branches, as the middle branch is pure photonic and no absorption take place, where the absorption is only for the polariton excitation part. Also here zero absorption at $\theta=0$, and maximum absorption at $\theta=\pi/2$, are obtained. Here, for the excitation and photon damping rates we used $\gamma/2\pi= 10^9\ Hz$ and $\Gamma_{ex}/2\pi= 10^8\ Hz$. The line width is wider for peaks and dips which are more photonic than excitation, where the line width for peaks and dips which are more photonic is dominated by the cavity line width, while for excitation dips and peaks the line width approaches the excitation line width.

\begin{figure}
\centerline{\epsfxsize=7.0cm \epsfbox{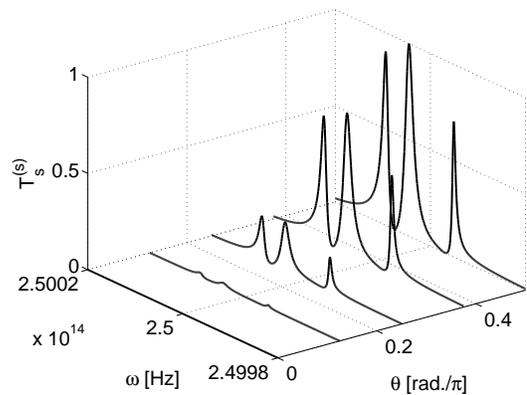}}
\caption{The $(s)$ polarized field transmission spectra, $T_s^{(s)}$, for different angles, $\theta$, at $k=5\times10^{-7}\ \AA^{-1}$ of $(s)$ polarized incident field.}
\end{figure}

\begin{figure}
\centerline{\epsfxsize=7.0cm \epsfbox{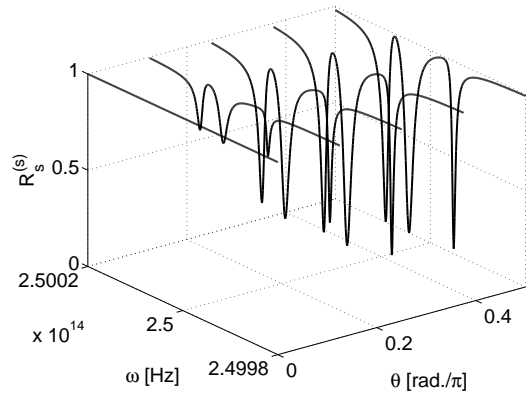}}
\caption{The $(s)$ polarized field reflection spectra, $R_s^{(s)}$, for different angles, $\theta$, at $k=5\times10^{-7}\ \AA^{-1}$ of $(s)$ polarized incident field.}
\end{figure}

\begin{figure}
\centerline{\epsfxsize=7.0cm \epsfbox{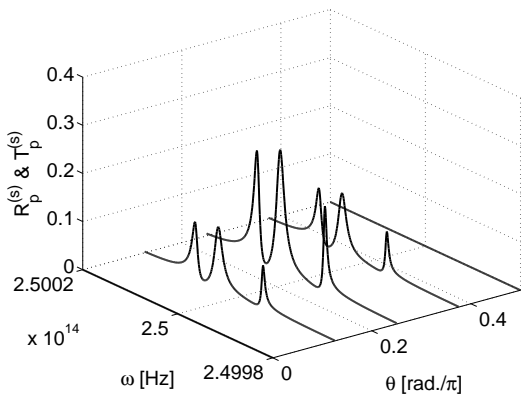}}
\caption{The $(p)$ polarized field transmission and reflection spectra, $T_p^{(s)}$ and $R_p^{(s)}$, for different angles, $\theta$, at $k=5\times10^{-7}\ \AA^{-1}$ of $(s)$ polarized incident field.}
\end{figure}

\begin{figure}
\centerline{\epsfxsize=7.0cm \epsfbox{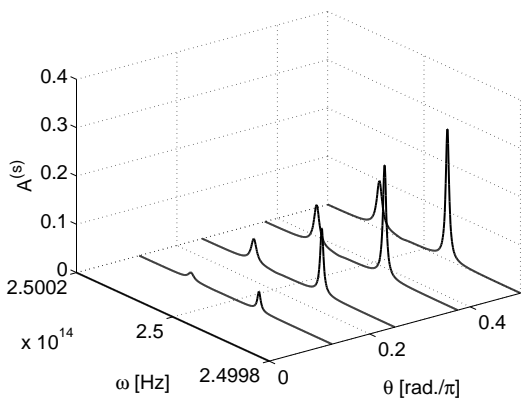}}
\caption{The absorption spectra, $A^{(s)}$, for different angles, $\theta$, at $k=5\times10^{-7}\ \AA^{-1}$ of $(s)$ polarized incident field.}
\end{figure}

The phase shift in the $(s)$ polarized transmitted field relative to the incident field, $\Delta\phi^{(s)t}_s$, as a function of frequency, at $k=5\times10^{-7}\ \AA^{-1}$ and for $\theta=\pi/4$, is plotted in figure (17). And the phase shift in the $(s)$ polarized reflected field, $\Delta\phi^{(s)r}_s$, is plotted in figure (18). In figure (19) we plot the phase shift in the transmitted and reflected $(p)$ polarized fields, $\Delta\phi^{(s)t}_p$ and $\Delta\phi^{(s)r}_p$. As the coupling through the mirrors is taken to be a real constant, $\gamma$, we neglect here phase shifts due to the cavity mirrors. In the general case they can be easily included.

\begin{figure}
\centerline{\epsfxsize=7.0cm \epsfbox{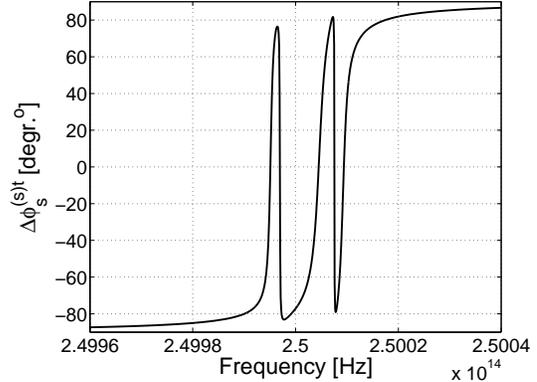}}
\caption{The phase shift in the $(s)$ polarized transmitted field, $\Delta\phi^{(s)t}_s$, at $k=5\times10^{-7}\ \AA^{-1}$ and for $\theta=\pi/4$ of $(s)$ polarized incident field.}
\end{figure}

\begin{figure}
\centerline{\epsfxsize=7.0cm \epsfbox{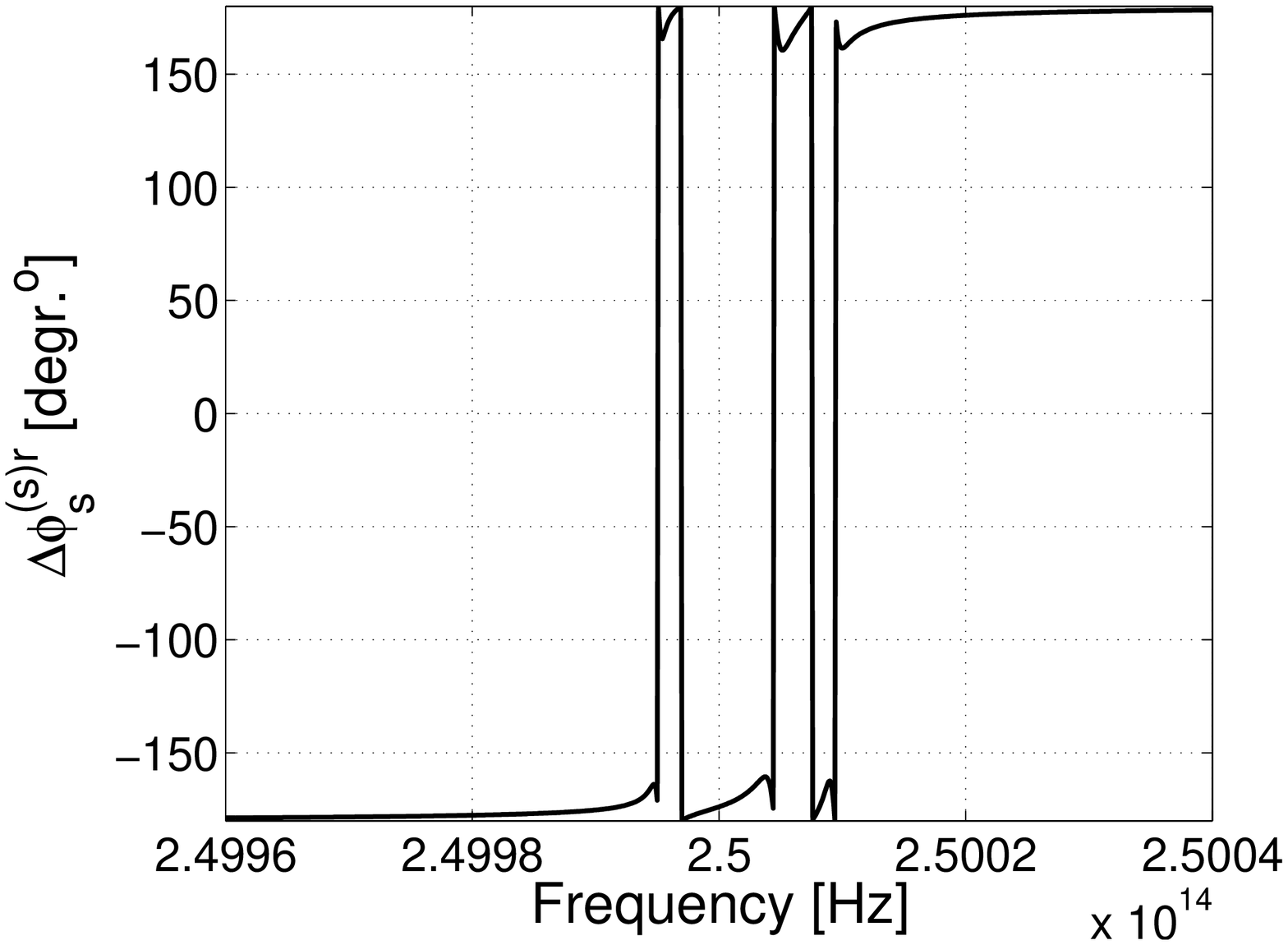}}
\caption{The phase shift in the $(s)$ polarized reflected field, $\Delta\phi^{(s)r}_s$, at $k=5\times10^{-7}\ \AA^{-1}$ and for $\theta=\pi/4$ of $(s)$ polarized incident field.}
\end{figure}

\begin{figure}
\centerline{\epsfxsize=7.0cm \epsfbox{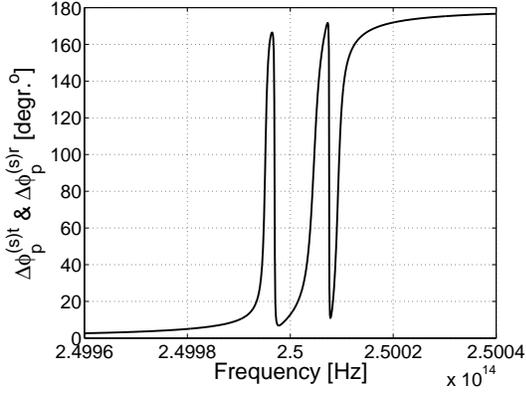}}
\caption{The phase shift in the $(p)$ polarized transmitted and reflected fields, $\Delta\phi^{(s)t}_p$ and $\Delta\phi^{(s)r}_p$, at $k=5\times10^{-7}\ \AA^{-1}$ and for $\theta=\pi/4$ of $(s)$ polarized incident field.}
\end{figure}

The $(s)$ polarized cavity mean photon number relative to the incident $(s)$ polarized photon number, in using $\sqrt{\gamma}\ \tilde{a}_{s}=\tilde{c}^{s}_{out}$, is given by
\begin{equation}
I^{(s)}_s=\frac{\langle\tilde{a}_{s}^{\dagger}\tilde{a}_{s}\rangle}{\langle\tilde{b}_{in}^{s\dagger}\tilde{b}^{s}_{in}\rangle}=\frac{\gamma\ |\Lambda_{ss}(1+\gamma\ \Lambda_{pp})-\gamma\ \Lambda_{sp}\Lambda_{ps}|^2}{|D|^2},
\end{equation}
and the $(p)$ polarized cavity mean photon number relative to the incident $(s)$ polarized photon number, in using $\sqrt{\gamma}\ \tilde{a}_{p}=\tilde{c}^{p}_{out}$, is given by
\begin{equation}
I^{(s)}_p=\frac{\langle\tilde{a}_{p}^{\dagger}\tilde{a}_{p}\rangle}{\langle\tilde{b}_{in}^{s\dagger}\tilde{b}^{s}_{in}\rangle}=\frac{\gamma\ |\Lambda_{ps}|^2}{|D|^2}.
\end{equation}
Note that $I^{(s)}_s=T^{(s)}_s/\gamma$, and $I^{(s)}_p=T^{(s)}_p/\gamma=R^{(s)}_p/\gamma$. Up to the division by $\gamma$, the plot of $I^{(s)}_s$ is as that of $T^{(s)}_s$ in figure (13), and the plot of $I^{(s)}_p$ is as that of $T^{(s)}_p$ and $R^{(s)}_p$ in figure (15). Hence the cavity mean photon number of both polarizations can be observed directly through the optical linear spectra.

Identical results are obtained for the case of $(p)$ polarized incident field only in exchanging $(s)$ and $(p)$ in all the previous equations. The incident field can be also in a superposition of both polarizations.

\section{Summary}

We investigated an anisotropic optical lattice, for the case of the Mott insulator phase with one atom per site. The anisotropy is induced by the atomic transition dipole of a fixed direction, which is fixed by the combination of the optical lattice laser polarization and external static fields. In the strong coupling regime, the cavity photon polarizations, of TE and TM modes, are coherently mixed with the electronic excitations to form two cavity polariton branches. As the superposition of the cavity photons of both polarizations has a component which is orthogonal to the atomic transition dipole, a third photonic branch is obtained that decouple to the electronic transitions.

The system eigenmodes are observed by linear optical spectra, which also proved the polarization mixing. For an incident field which is TE polarized, we get TE and TM transmission and reflection spectra, where the TM spectra is induced by the anisotropic optical lattice. The absorption spectrum of the cavity medium is calculated in including the atom excited state life time phenomenologically. Furthermore, the eigenmodes and the polarization mixing can be observed through the phase shift of the transmitted and reflected fields relative to the incident field.

Polarization mixing is of big importance for electro-optics devices and quantum information, and can be used as an observation tool of many properties of anisotropic optical lattices. The system can serve as a linear optical switch, as the transmitted and reflected field intensities and phase shifts are a function of the angle between the incident field and the transition dipole, they can be controlled in changing the angle. Moreover, the results allow us to fix the mean photon number in the cavity, for both polarizations, in fixing the intensity of the incident field.

\begin{acknowledgments}
The work was supported by the Austrian Science Fund (FWF), through the Lise-Meitner Program (M977).  
\end{acknowledgments}

\end{document}